\documentclass[twocolumn,showpacs,amsmath,nofootinbib,amssymb,superscriptaddress]{revtex4}

\usepackage{graphicx}
\usepackage{dcolumn}
\usepackage{bm}

\begin{document}

\title{Energy conditions in $f(R)$--gravity}

\author{J. Santos}\email{janilo@dfte.ufrn.br}
\affiliation{Universidade Federal do Rio Grande do Norte, 
Departamento de F\'{\i}sica C.P. 1641, 59072-970 Natal -- RN, Brasil}
\affiliation{Departamento de Astronomia, Observat\'orio Nacional, 
20921-400 Rio de Janeiro -- RJ, Brasil}
\affiliation{Centro Brasileiro de Pesquisas F\'{\i}sicas, 
Rua Dr.\ Xavier Sigaud 150,\ 22290-180 Rio de Janeiro -- RJ, Brasil}

\author{J.S. Alcaniz}\email{alcaniz@on.br}
\affiliation{Departamento de Astronomia, Observat\'orio Nacional, 
20921-400 Rio de Janeiro -- RJ, Brasil}

\author{M.J. Rebou\c{c}as}\email{reboucas@cbpf.br}
\affiliation{Centro Brasileiro de Pesquisas F\'{\i}sicas, 
Rua Dr.\ Xavier Sigaud 150,\ 22290-180 Rio de Janeiro -- RJ, Brasil}

\author{F.C. Carvalho}\email{fabiocc@on.br}
\affiliation{Departamento de Astronomia, Observat\'orio Nacional, 
20921-400 Rio de Janeiro -- RJ, Brasil}

\date{\today}

\begin{abstract}
In order to shed some light on the current discussion about $f(R)$--gravity theories we derive and discuss the bounds imposed by the energy conditions on a general $f(R)$ functional form. The null and strong energy conditions in this framework are derived from the Raychaudhuri's equation along with the requirement that gravity is attractive,  whereas the weak and dominant energy conditions are stated from a comparison with the energy conditions that can be obtained in a direct approach via an effective energy-momentum tensor for $f(R)$--gravity. As a concrete application of the energy conditions to locally homogeneous and isotropic $f(R)$--cosmology, the recent estimated values of the deceleration and jerk parameters are used to examine the bounds from 
the weak energy condition on the parameters of two families of $f(R)$--gravity theories. 
\end{abstract}

\maketitle

\section{Introduction}

The observed late-time acceleration of the Universe poses one of the greatest challenges theoretical physics has ever faced. In principle, this phenomenon may be the result of unknown physical processes involving either modifications of gravitation theory or the existence of new fields in high energy physics. Although the latter route is most commonly used, following the former, an attractive and complementary approach to this problem, known as $f(R)$--gravity~\cite{Kerner}, examines the possibility of modifying Einstein's general relativity (GR) by adding terms proportional to powers of the Ricci scalar $R$ to the Einstein-Hilbert Lagrangian\footnote{Another interesting approach is related to the possible existence of extra dimensions,  an idea that  links cosmic acceleration with the hierarchy problem in high energy physics, and gives rise to the so-called brane-world cosmology~\cite{bw}.}.

Recently, several different functional forms for $f(R)$ have been suggested  in the literature (see, e.g., Ref.~\cite{frgravity}). These different $f(R)$--gravity  theories have also been discussed in different contexts as, e.g.,  in the issues related to the stability conditions~\cite{Dolgov}, inflationary epoch~\cite{barrow}, compatibility with solar-system tests and galactic data~\cite{Chiba}, the late-time cosmological evolution~ \cite{Amendola}, among others. In the cosmological context, although these theories provide an alternative way to explain the cosmic speed up without dark energy, the freedom in building different functional forms of $f(R)$ gives rise to the problem of how to constrain from theoretical and/or observational aspects these many possible $f(R)$-gravities. Recently, this possibility has been explored by testing the cosmological viability of some specific forms of $f(R)$ (see, e.g., \cite{Amendola}). 

Additional constraints to $f(R)$ theories may also come by  imposing the  so-called energy conditions~\cite{Kung,Bergliaffa} (for a pedagogical review, see Ref.~\cite{Carroll}).  As is well known, these conditions were used in different contexts to derive general results that hold for a variety of situations. Thus, for example, the Hawking-Penrose singularity theorems invoke the weak (WEC) and strong (SEC) energy conditions, whereas the proof of the second law of black hole thermodynamics requires the null energy condition (NEC). More recently, several authors have used the classical energy conditions of GR to investigate some cosmological issues, such as the phantom fields potentials~\cite{Santos}, expansion history of the universe~\cite{Nilza}, as well as evolution of the deceleration  parameter and their confront with supernovae observations~\cite{Gong}.

An important aspect worth emphasizing is that the energy conditions were initially formulated in the context of GR~\cite{Hawking}. In other words, this amounts to saying that one has to be cautious when using them in the more general framework, such as the $f(R)$--gravity. On the other hand, we note that $f(R)$--gravity theories share an interesting property: starting from the Jordan frame, where gravity is described only by the metric tensor, and making a suitable conformal transformation on the metric, it can be shown that any $f(R)$ theory is mathematically equivalent to Einstein's gravity with 
a minimally coupled scalar field~\cite{Barrow-Cotsakis-Magnano}. 

Thus, in principle, one could think of ``translating" the energy conditions directly from GR and impose them on the new effective pressure and effective energy density defined in the Jordan frame. To test the viability of such a procedure, in this paper we derive the energy 
conditions for general  $f(R)$--gravity by using the physical ultimate origin of the NEC and SEC, which is the Raychaudhuri equation along with the 
requirement that gravity is attractive. We find that the NEC and the SEC for $f(R)$, although similar, are different from that of Einstein's gravity. The resulting inequalities are then compared with what would be obtained by translating these energy conditions directly from the effective energy-momentum tensor for $f(R)$--gravity in the Jordan frame. There emerges from this comparison a natural statement of the weak and dominant energy conditions in the context of the $f(R)$--gravity theories. As a concrete application of the energy conditions for locally homogeneous and isotropic $f(R)$--cosmology, we use the recent estimated values of the deceleration and jerk parameters, to examine the bounds from the WEC on the parameters of two families of $f(R)$--gravity.

\section{Energy Conditions}

\subsection{General Relativity}

Much of the technique we will use to derive the energy conditions for $f(R)$ theories can be borrowed from the approach to the NEC and SEC in the context of GR. Therefore, for the sake of completeness, we shall first briefly review the approach to these energy conditions in GR. The ultimate origin of these energy conditions is the Raychaudhuri equation along with the requirement that gravity is attractive. To make clear this point, let $u^{\mu}$ be the tangent vector field to a congruence of timelike geodesics in a spacetime manifold endowed
with a metric $g_{\mu \nu}\,$. 
Raychaudhuri's equation (see Ref.~~\cite{Kar-Dadhich} for  recent reviews) then reads
\begin{equation}  \label{Raych}
\frac{d\theta}{d\tau}= - \frac{1}{3}\,\theta^2 -
\sigma_{\mu\nu}\sigma^{\mu\nu} + \omega_{\mu\nu}\omega^{\mu\nu} 
- R_{\mu\nu}u^{\mu}u^{\nu} \;,
\end{equation}
where  $R_{\mu\nu}$  is the Ricci tensor, and $\theta\,$, $\sigma^{\mu\nu}$ and $\omega_{\mu\nu}$ are, respectively, the  expansion, shear and rotation associated to the congruence defined by the vector field $u^{\mu}$. This equation is a purely geometric statement, and as such it makes no reference to any gravitational field equations. However, since the GR field equations relate $R_{\mu \nu}$ to the energy-momentum tensor $T_{\mu \nu}$, the combination of Einstein's and Raychaudhy's equations can be used to restrict energy-momentum tensors on physical grounds. Indeed, since  the shear is a ``spatial" tensor one has 
$\sigma^2 \equiv \sigma_{\mu\nu}\sigma^{\mu\nu}\geq 0$, thus from Raychaudhury's equation it is clear for any hypersurface orthogonal congruences ($\omega_{\mu\nu} \equiv 0$) the condition for attractive gravity (convergence of timelike geodesics or geodesic focusing)
reduces  to $R_{\mu\nu}u^{\mu}u^{\nu}\geq 0$, which by virtue of Einstein's equation implies
\begin{equation} \label{StrongEC}
R_{\mu\nu}\, u^{\mu}u^{\nu}=\left( T_{\mu\nu} - \frac{T}{2}g_{\mu\nu} \right)\,
u^{\mu}u^{\nu}\geq 0\,,
\end{equation}
where $T_{\mu\nu}$ is the energy-momentum tensor, $T$ is its trace, and
where we have used units such that $8\pi G=c=1$, which we shall adopt 
hereafter. Equation~(\ref{StrongEC}) is nothing but the SEC stated in a 
coordinate-invariant way in terms of $T_{\mu\nu}$ and vector fields of 
fixed (timelike) character~\cite{Carroll}. In this way, the SEC
in the GR context encodes the fact that gravity is attractive.
In particular, note that for a perfect fluid of density $\rho$ and 
pressure $p\,$ 
\begin{equation} \label{perf-fluid}
T_{\mu\nu}= (\rho + p)\,u_{\mu}u_{\nu} - p\,g_{\mu\nu}\;, 
\end{equation}
and the restriction given by Eq.~(\ref{StrongEC}) takes the familiar form 
for the SEC, i.e, $\rho + 3p \geq 0$.

The evolution equation for the expansion of a congruence of null geodesics
defined by a vector field $k^\mu$ has the same form of the 
Raychaudhury  equation, but with a factor $1/2$ rather than $1/3$, and 
with $-R_{\mu\nu}k^{\mu}k^{\nu}$ 
as the last term (see Ref.~\cite{Carroll} for more details). In 
this case, the condition for the convergence (geodesic focusing) of 
hypersurface orthogonal ($\widehat{\omega_{\mu\nu}} \equiv 0$) congruences 
of null geodesics  along with Einsteins's equation implies 
\begin{equation} \label{NullEC}
R_{\mu\nu}k^{\mu}k^{\nu} = T_{\mu\nu}k^{\mu}k^{\nu}\geq 0\,,
\end{equation}
which is the NEC written in a coordinate-invariant way~\cite{Carroll}.
Thus, in GR the NEC ultimately encodes the null geodesic focusing due 
to the gravitational attraction. 
We note that for the energy-momentum tensor of a perfect fluid 
[Eq.~(\ref{perf-fluid})] it reduces to well-known form of the NEC, 
i.e., $\rho + p \geq 0$.

\subsection{$f(R)$--Gravity}

Since the Raychaudhuri equation and its null-geodesics counterpart 
hold for any geometrical theory of gravitation, we will keep the 
above physical motivation for the SEC and NEC in the $f(R)$--gravity 
context. To this end, we shall use the above Raychaudhuri's geometric 
relations [see Eqs.~(\ref{StrongEC}) and (\ref{NullEC})] along with 
the attractive character of gravitational interaction to derive
these energy conditions in the context of $f(R)$--gravity theories. 

We begin by recalling that the action that defines an 
$f(R)$--gravity is given by
\begin{equation}
 S=\int d^4x\sqrt{-g}\,f(R) + S_m
\end{equation}
where $g$ is the trace of the metric $g_{\mu \nu}$, $R$ is the Ricci scalar 
and $S_m$ the standard action for the matter fields. Varying this action with 
respect to the metric we obtain the field equations
\begin{equation}  \label{field_eq}
f'R_{\mu\nu} - \frac{f}{2}g_{\mu\nu} - \left(\nabla_{\mu}\nabla_{\nu}-
g_{\mu\nu}\,\Box \,\right)f' = T_{\mu\nu}\,,
\end{equation}
where a prime denotes differentiation with respect to $R$ and 
$\Box \equiv g^{\alpha \beta}\,\nabla_{\alpha}\nabla_{\beta}\,$.
In order to use an approach to the NEC and SEC similar to that in GR context, 
we note that Eq.~(\ref{field_eq}) can be rewritten as 
\begin{equation} \label{Ricci-f}
R_{\mu\nu} = {\cal{T}}_{\mu\nu} -\frac{{\cal{T}}}{2}g_{\mu\nu}\;,
\end{equation}
where \vspace{-4mm}
\begin{eqnarray} {\cal{T}} &=& \frac{1}{f'}(T +f-Rf'-3\Box f') \;,\nonumber\\ 
{\cal{T}}_{\mu\nu}&=&\frac{1}{f'}\left[ T_{\mu\nu} + (\nabla_{\mu}\nabla_{\nu}-
g_{\mu\nu}\Box)f'   \right] \nonumber \,,
\end{eqnarray}
and $\,T = g^{\mu \nu}T_{\mu \nu}\,$.

Now, for the homogeneous and isotropic Friedmann-Lema\^{\i}tre-Robertson-Walker 
(FLRW) metric with scale factor $a(t)$, the relations $R_{\mu\nu}k^{\mu}k^{\nu}\geq 0$  
and  $R_{\mu\nu}u^{\mu}u^{\nu}\geq 0$ [see Eqs.~(\ref{StrongEC}) and (\ref{NullEC})] 
along with Eq.~(\ref{Ricci-f}) lead, for a perfect fluid $T_{\mu\nu}\,$, to the 
following inequalities:
\begin{equation}  \label{NEC-f}
\rho + p  \geq 0\;, \quad \mbox{and} \quad  \left( \ddot{R}-\dot{R}\,H \right)f'' + \dot{R}^2f''' \geq 0\,,
\end{equation}
\begin{equation}  \label{SEC-f}
\rho + 3p  - f + Rf' + 3\left( \ddot{R}+\dot{R}\,H\right)f'' + 3\dot{R}^2f''' \geq 0\,,
\end{equation}
where a dot denotes derivative with respect to time, $H=\dot{a}/a$ is the Hubble parameter, and we have made the assumption that $f'>0$, which ensures the regularity of the conformal rescaling~\cite{Barrow-Cotsakis-Magnano}. Thus, the inequalities~(\ref{NEC-f}) and~(\ref{SEC-f}) are, respectively, the 
NEC and SEC requirements in the context of $f(R)$--gravity for a perfect fluid FLRW model. We note that the well-known forms for the NEC ($\rho + p  \geq 0$) and 
SEC ($\rho + 3p  \geq 0$) in the context GR can be recovered as a particular case of these energy conditions in $f(R)$ theories for $f=R$, as one would have expected.

An important point to derive the week and dominant energy conditions (WEC and DEC, respectively)  in $f(R)$--gravity is the fact that the above NEC and SEC [Eqs.(\ref{NEC-f}) and (\ref{SEC-f})] can also be recovered as an extension of these conditions in GR. In fact, in $f(R)$--gravity theories one can 
define an effective energy-momentum tensor as (see, e.g., Ref.~\cite{Hwang})
\begin{equation}  \label{Teff} 
T^{\mbox{eff}}_{\mu\nu}= \frac{1}{f'}\left[ T_{\mu\nu} 
+ \frac{1}{2}(f - Rf')g_{\mu\nu} 
+ (\nabla_{\mu}\nabla_{\nu} - g_{\mu\nu}\Box )f' \right] \,,
\end{equation} 
from which one defines an effective energy density and pressure by
\begin{equation}\label{rho_eff}
 \rho_{\mbox{eff}} =  \frac{1}{f'}\left[ \rho 
 + \frac{1}{2}(f - Rf') - 3\dot{R}Hf''\right] \;, 
\end{equation}
and 
\begin{equation} \label{p_eff}
p_{\mbox{eff}}= \frac{1}{f'} \left[ p - \frac{1}{2}(f - Rf') 
+ (\ddot{R}+2\dot{R}H)f'' +  
\dot{R}^2f''' \right] \,,
\end{equation}
which in turn make apparent that the NEC and WEC given by Eqs.~(\ref{NEC-f}) and~(\ref{SEC-f}) can be obtained in a similar way as that in GR context. 
Thus, following and extending the GR approach to WEC and DEC, we have that in $f(R)$--gravity theories, in addition to Eq.~(\ref{NEC-f}), the WEC requires that
\begin{equation}  \label{WEC-f}
\rho + \frac{1}{2}(f - Rf') - 3\dot{R}Hf'' \geq 0 \,,
\end{equation}
whereas the DEC fulfillment, besides the inequalities~(\ref{NEC-f}) 
and (\ref{WEC-f}), reads 
\begin{equation}  \label{DEC-f}
\rho - p +  f - Rf' - (\ddot{R}+ 5\dot{R}H)f'' - \dot{R}^2f''' \geq 0 \,.
\end{equation} 
As one may easily check, for $f=R$,  Eqs.~(\ref{WEC-f}) and~(\ref{DEC-f}) give $\rho \geq 0$ and $\rho - p\geq 0$, whose combination with Eqs~(\ref{NEC-f}) give, respectively, the well-known forms of the WEC and DEC in GR (see, e.g., Ref.~\cite{Nilza}). 
We also note that according to Ref.~\cite{Barrow-Ottewill} if the energy momentum
tensor is trace-free, then any homogeneous and isotropic solution of GR 
is also a particular solution of a $f(R)$--theory provided $f(0)=0\,$, 
and $f'(0) \neq 0 $. Thus, it is clear from the above  
Eqs.~(\ref{NEC-f}), (\ref{SEC-f}), (\ref{WEC-f}) and (\ref{DEC-f}) 
that in these cases the GR energy conditions are recovered.  

For the sake of completeness, we note that rather than the geodesic focusing the 
so-called sudden future singularity (SFS) is another type of singularity 
that arises in the context of perfect fluid FLRW cosmology~\cite{Barrow2004,Barrow2004a,Stefancic,Nojiri,Dabrowski}. It comes 
about in the absence of an precise equation of state and is a singularity of 
pressure $p$ only, with finite energy density $\rho$. In the context of GR, the 
SFS can occur with no violation of the SEC and WEC but with violation of the DEC
~\cite{Barrow2004,Barrow2004a,Stefancic,Nojiri,Dabrowski}. 
The nature of SFS in terms of geodesic completeness has been discussed in 
Ref.~\cite{Lazkoz04}.
A discussion of the SFS in the context of the $f(R)$--gravity, however, is beyond 
the scope of the present paper. In this regards, we refer the readers 
to Ref.~\cite{Barrow2004a} (see also Ref.~\cite{Barrow-Tsagas2005}).

\vspace{0.3cm}
\section{Constraining $f(R)$ Theories}

The energy-condition inequalities~(\ref{NEC-f}), (\ref{SEC-f}),  (\ref{WEC-f}) and~(\ref{DEC-f}) can also be used to place bounds on a given $f(R)$ in the context of FLRW models. To investigate such bounds, we first note that the Ricci scalar and its derivatives for a spatially flat FLRW geometry can be expressed in terms of the deceleration ($q$), jerk ($j$) and snap ($s$) parameters, i.e, 
\begin{subequations}
\label{eq:whole}
\begin{eqnarray}
R &=&-6H^2(1-q)\;, \\
\dot{R}&=&-6H^3(j-q-2)\;, \\
\ddot{R}&=&-6H^4(s+q^2+8q+6)\;, 
\end{eqnarray}
\end{subequations}
where~\cite{Visser}
\begin{equation}
q=-\frac{1}{H^2}\frac{\ddot{a}}{a}\;, \qquad j=\frac{1}{H^3}\frac{\dddot{a}}{a}\;, 
\quad {\rm{and}} \quad s=\frac{1}{H^4}\frac{\ddddot{a}}{a}\;.
\end{equation}
In terms of the present-day values for the above parameters, Eqs.~(\ref{NEC-f}), (\ref{SEC-f}),  (\ref{WEC-f}) and (\ref{DEC-f}) can be rewritten as
\begin{widetext}\label{today_EC}
\begin{subequations}
\begin{eqnarray} 
(\mbox{\bf NEC}) &  \rho_0 + p_0 \geq 0 \quad \mbox{and} \quad
   -[s_0-j_0+(q_0+1)(q_0+8)]f''_0 
+  6[H_0(j_0 -q_0 -2)]^2f'''_0 \geq 0\,,  \\
\nonumber \\
(\mbox{\bf WEC}) & \quad \quad \quad  2 \rho_{0} + f_0 + 6 H^2_0(1-q_0)f'_0 
+ 36 H^4_0(j_0-q_0-2)f''_0   \geq 0\,, \label{wec0} \\
\nonumber \\
(\mbox{\bf SEC})  &\;\;   \rho_0 +3 p_0 + f_0 - 6H^2_0(1-q_0)f'_0 - 
+6 H^4_0(s_0+j_0+q^2_0 + 7q_0+4)f''_0  + 3[6H^3_0(j_0 - q_0 -2)]^2f'''_0  \geq 0\,,\\
\nonumber \\
(\mbox{\bf DEC}) &\quad   \rho_0 - p_0 + 6H^2_0(1-q_0)f'_0 - 6H^4_0[s_0 
+ (q_0-1)(q_0+4) +5j_0]f''_0  - [6H^3_0(j_0-q_0-2)]^2f'''_0  \geq 0\,.
\end{eqnarray}
\end{subequations}
\end{widetext}

\subsection{$f(R) = R + \alpha R^n$}

To exemplify how the above conditions can be used to place bounds on $f(R)$ theories, we first note that, apart from the WEC [Eq.~(\ref{wec0})], all above inequalities depend on the current value of the snap parameter $s_0$. Therefore, since no reliable measurement of this parameter has been reported hitherto, in what follows we shall focus on the WEC requirement in the confrontation of the energy condition bounds in $f(R)$---gravity with observational data.

As a first concrete example, we shall consider the family of  theories with $f(R)$ of the form
\begin{equation}  \label{f_R} 
f(R) = R + \alpha R^n\;,
\end{equation}  
where $n$ is an integer and $\alpha$ is a constant that can assume positive or negative values. For this class of theories we note that the constraints from WEC depend only on observational values $q_0$ and $j_0$, besides $n$ and $\alpha$ that specify a particular theory. The inequality for WEC fulfillment condition~(\ref{wec0}) can be written in terms of these parameters as
\begin{equation}  \label{signal_1}
 \alpha (-1)^n(An^2 -(A+1)n + 1)\geq 0\,,
\end{equation}
where $A=(j_0 - q_0 -2)/(1-q_0)^2$. In what follows we consider $q_0=-0.81\pm$ 0.14 and $j_0=2.16^{+0.81}_{-0.75}$, as given in Ref.~\cite{Rapetti}. The roots of the quadratic function are ($1,1/A$), so it has positive values for $1\geq n \geq 3.377$ and negative values for $1 < n < 3.377$. 

Inequation~(\ref{signal_1}) makes apparent that the observance of the WEC clearly depend upon the sign of $\alpha$ and the values of $n$. In what follows we consider the two possible signs for $\alpha$ along with the requirement that $f'(R)>0\,$ for all $\,R$, which reduces to 
\begin{equation}  \label{signal_2}
 \alpha (-1)^n\,{n\,(3.3H_0)^{2n-2}} < 1
\end{equation}  
for $f(R)$ given by~(\ref{f_R}). Thus, for the family of $f(R)$ theories given by Eq.~(\ref{f_R}) the WEC is obeyed in the following cases: 
\begin{itemize}

\item[{\bf (i)}] $\alpha > 0\,$. Here the allowed values for $n$ can be grouped in following sets: $n=\{3,1,-2,-4,-6,\ldots\}$ and $n=\{4,6,8,\ldots\}$, with $0<\alpha < [n(3.3H_0)^{2n-2}]^{-1}$ for this last set.

\item[{\bf (ii)}] $\alpha < 0\,$. In this case the allowed sets are $n=\{2,-1,-3,-5,\ldots\}$ and $n=\{1, 5,7,9,\ldots\}$, with $\alpha$ within the range $-[n(3.3H_0)^{2n-2}]^{-1}<\alpha <0$ for this last set.
.
\end{itemize}

As a concrete application, it is clear from the above analysis that the so-called quadratic gravity, i.e. $R + \alpha R^2$ gravity (see, e.g., Ref.~\cite{Kung}) obey the WEC only if $\alpha$ is negative. Another interesting example is the $f=R-\mu^4/R$ gravity theory (see Ref.~\cite{Turner}), which correspond to WEC fulfillment 
case $\alpha <0,\;n=-1$. The closely related  $f=R-\mu^6/R^2$ theories, however,  violates the WEC, since it corresponds to $\alpha <0$ and $n=-2$. We also note that the violation of WEC by a specific member of a class of $f(R)$--gravity theories clearly lead to the breakdown of the DEC. 

\vspace{0.1cm}

\subsection{$f(R)=\alpha R^n$}

This type of $f(R)$--gravity theories has been proposed in the investigation  of galactic environments~\cite{Sobouti,Capo}, gravitational waves and lensing effects~\cite{Mendoza}, as well as FLRW dynamics via phase space analysis~\cite{Carloni,Timothy_John}. For this class of theories the WEC constraints are again given by Eq.~(\ref{signal_1}), but the requirement that $f'(R)>0$ (for all $R$) now takes the form 
\begin{equation}  \label{signal_3}
  \alpha (-1)^n\, n\,(3.3H_0)^{2n-2} < 0\,.
\end{equation} 
As before we group this class of gravity theories in two cases depending  on the sign of $\alpha$. We find that, for integers values of $n$,  the WEC is obeyed in the following cases:
\begin{itemize}
\item[{\bf (i)}] $\alpha > 0\,$. The allowed values for $n$ are  $n=\{3,1,-2,-4,-6,\ldots\}$. 

\item[{\bf (ii)}] $\alpha < 0$. In this case we have the set 
$n=\{2,-1,-3,-5,\ldots\}$.

\end{itemize}
Finally, we note that the WEC-fulfillment case ${\bf (i)}$  may accommodate the observational value $n=3.5$ obtained by fitting the Supernovae Ia Hubble diagram with this type of power-law gravity~\cite{Capo}.
In Ref.~\cite{Timothy_John} the authors found that this type of
theories admit simple exact solutions for FLRW models and that a
stable matter-dominated period of evolution requires $n>1$ or $n<3/4$. 
Clearly, the first constraint do not violate the WEC, while the
second do.

\section{Final Remarks}

$f(R)$--gravity provides an alternative way to explain the current cosmic acceleration with no need of invoking either the existence of an extra spatial dimension or an exotic component of dark energy. However, the arbitrariness in the choice of different functional forms of $f(R)$  gives rise to the problem of how to constrain the many possible $f(R)$--gravity theories on physical grounds.

In this paper we have shed some light on this question by discussing some constraints on general $f(R)$--gravity from the  so-called energy conditions. Starting from the Raychaudhuri's equation along with the  requirement that gravity is attractive, we have derived the null and strong energy conditions in the framework of $f(R)$--gravity  and shown that, although similar, they differ from those derived in the context of general relativity. By comparing the NEC and SEC inequalities [Eqs. (\ref{NEC-f}) and (\ref{SEC-f})] with what would be obtained by translating these energy conditions directly from the effective energy-momentum tensor for $f(R)$--gravity in the Jordan frame, we also have obtained the general expressions for the weak and dominant energy conditions [Eqs. (\ref{WEC-f}) and (\ref{DEC-f})]. As a concrete example of how these energy conditions requirements may constrain $f(R)$--gravity theories, we have discussed the WEC bounds on two different $f(R)$ classes, i.e., $f(R) = R + \alpha R^n$ and $f(R)=\alpha R^n$. An interesting outcome of this analysis is that the so-called quadratic gravity ($R + \alpha R^2$), originally discussed in the context of primordial inflation, violates the WEC requirement for positive values of $\alpha\,$.

Finally, we emphasized that although the energy conditions in $f(R)$--gravity discussed 
in this article has well-founded physical motivation (Raychadhuri's equation along with 
the attractive character of gravity) the question as to whether they should be 
applied to any solution of $f(R)$--gravity theories is an open question, which  is 
ultimately related to the confrontation between theory and observations. We recall,
however, that this confrontation in GR context indicate that all energy 
conditions seem to have been violated at some point of the recent past of cosmic 
evolution~\cite{Nilza}.

\begin{acknowledgments}
J. Santos acknowledges financial support from PCI-CBPF/MCT and PRONEX (CNPq/FAPERN), 
and the kind hospitality of Observat\'orio Nacional/MCT. F.C. Carvalho acknowledges 
financial support from PCI-ON/MCT. J.S.A. and M.J.R. thank CNPq for the grants under 
which this work was carried out. JSA is also supported by FAPERJ No. E-26/171.251/2004.
\end{acknowledgments}



\begin{thebibliography}{99}

\bibitem{Kerner} R. Kerner, Gen. Relativ. Gravit. {\bf 14}, (1982) 453.

\bibitem{bw}  L. Randall and R. Sundrum, Phys. Rev. Lett. {\bf{83}}, (1999), 3370; {\bf{83}}, (1999), 4690;  G. Dvali, G. Gabadadze, and M. Porrati, Phys. Lett. B {\bf{485}}, (2000), 208;    C. Deffayet {\it et al.}, Phys. Rev. D {\bf{66}}, (2002), 024019;  V. Sahni and Y. Shtanov, Int. J. Mod. Phys. D {\bf{11}}, (2002), 1515;  J. Cosmol. Astropart. Phys. {\bf{11}}, (2003), 014; J. S. Alcaniz, Phys. Rev. D {\bf{65}}, (2002), 123514. arXiv:astro-ph/0202492;  D. Jain, A. Dev, and J. S. Alcaniz, Phys. Rev. D {\bf{66}}, (2002), 083511. arXiv:astro-ph/0206224; M. D. Maia, E. M. Monte, J. M. F. Maia, and J. S. Alcaniz, Class. Quantum Grav. {\bf{22}}, (2005), 1623. arXiv:astro-ph/0403072;  A. Lue, Phys. Rep. {\bf{423}}, (2006), 1.

\bibitem{frgravity} S. Nojiri and S.D. Odintsov, Phys. Rev. D {\bf 68}, (2003) 123512; Phys. Lett. B {\bf 576}, (2003) 5;  Gen. Relativ. Gravit. {\bf 36}, (2004) 1765; S.M. Carroll, V. Duvvuri, M. Trodden and M.S. Turner, Phys. Rev. D {\bf 70}, (2004) 043528;  X.H. Meng and P. Wang, Phys. Lett. B {\bf 584}, (2004) 1;  S. Capozziello, V.F. Cardone and A. Troisi, Phys. Rev. D {\bf 71}, (2005) 043503;  A. Cruz-Dombriz and A. Dobado, Phys. Rev. D {\bf 74}, (2006) 087501; S. Nojiri and S.D. Odintsov, Phys. Rev. D {\bf 74}, (2006) 086005; S.K. Srivastava, Phys. Lett. B {\bf 643}, (2006) 1; S. Nojiri and S.D. Odintsov, Int. J. Geom. Meth. Mod. Phys. {\bf 4}, (2007) 115; S. Capozziello and M. Francaviglia, arXiv:0706.1146v1 [astro-ph]; S.A. Appleby and R. Battye, arXiv: 0705.3199 [astro-ph]; 
A.A. Starobinsky, arXiv:0707.2041 [astro-ph]; 
D.A. Easson, Int.\ J.\ Mod.\ Phys.\  A {\bf 19}, 5343 (2004);
J.A.R. Cembranos, Phys.Rev. D {\bf 73},  064029 (2006);
O. Bertolami, C.G. Boehmer, T. Harko, F.S.N. Lobo, Phys. Rev. D {\bf 75}, 104016 (2007).

\bibitem{Dolgov} A.D. Dolgov and M. Kawasaki, Phys. Lett. B {\bf 573}, (2003) 1; V. Faraoni and S. Nadeau, Phys. Rev. D {\bf 72}, (2005) 124005; V. Faraoni, Phys. Rev. D {\bf 74}, (2006) 104017; C.G. Boehmer, L. Hollenstein and F.S.N. Lobo, arXiv:0706.1663 [gr-qc].

\bibitem{barrow} J.D. Barrow and S. Hervik, Phys. Rev. D {\bf 73}, (2006) 023007

\bibitem{Chiba} T. Chiba, Phys. Lett B {\bf 575}, (2003) 1; G.J. Olmo, Phys. Rev. Lett. {\bf 95}, (2005) 261102; G.J. Olmo, Phys. Rev. D {\bf 72}, (2005) 083505; O. Mena, J. Santiago and J. Weller, Phys. Rev. Lett. {\bf 96}, (2006) 041103;  T. Chiba, T.L. Smith and A.L. Erickcek, Phys. Rev. D {\bf{75}},  (2007) 124014; B. Li and J.D. Barrow, Phys. Rev. D {\bf 75}, (2007) 084010; I. Navarro and K. Van Acoleyen, JCAP 0702 (2007) 022;  P. Zhang, astro-ph/0701662v1; 
L. Amendola and S. Tsujikawa, arXiv:0705.0396v1 [astro-ph] (2007).

\bibitem{Amendola} S. Capozziello, S. Nojiri, S.D. Odintsov and A. Troisi, Phys. Lett. B {\bf 639}, (2006) 135; L. Amendola, D. Polarski and S. Tsujikawa, astro-ph/0605384 (2006); A. W. Brookfield, C. van de Bruck and L.M.H. Hall, 
Phys. Rev. D {\bf 74}, (2006) 064028;  L. Amendola, R. Gannouji, D. Polarski and S. Tsujikawa, Phys. Rev. D {\bf 75}, (2007) 083504; M. Fairbairn and S. Rydbeck, arXiv:astro-ph/0701900v2 (2007); L. Amendola, D. Polarski and S. Tsujikawa, Phys. Rev. Lett. {\bf 98}, (2007) 131302;
S. Fay, S. Nesseris, L. Perivolaropoulos, arXiv:gr-qc/0703006v3.

\bibitem{Kung} J.H. Kung, Phys. Rev. D {\bf 52}, (1995) 6922; Phys. Rev. D {\bf 53}, (1996) 3017. 

\bibitem{Bergliaffa} S.E. Perez Bergliaffa, Phys. Lett. B {\bf 642}, (2006) 311.

\bibitem{Carroll} S. Carroll, {\em Spacetime and Geometry: An Introduction to General Relativity}, (Addison Wesley, New York, 2004).

\bibitem{Santos} J. Santos and J.S. Alcaniz, Phys. Lett. B {\bf 619}, (2005) 11. arXiv:astro-ph/0502031.

\bibitem{Nilza} M. Visser, Science {\bf 276}, (1997) 88; Phys. Rev. D {\bf 56}, (1997) 7578; 
J. Santos, J.S. Alcaniz and M.J. Rebou\c{c}as, Phys. Rev. D {\bf 74}, (2006) 067301. arXiv:astro-ph/0608031;  
J. Santos, J.S. Alcaniz, N. Pires and M.J. Rebou\c{c}as, Phys. Rev. D {\bf 75}, (2007) 083523. arXiv:astro-ph/0702728; A.A. Sen and R.J. Scherrer, arXiv:astro-ph/0703416; J. Santos, J.S. Alcaniz, M.J. Rebou\c{c}as and N. Pires, arXiv: 0706.1779 [astro-ph] (2007).
 
\bibitem{Gong} Y.G. Gong, A. Wang, Q. Wu and Y.Z. Zhang, arXiv:astro-ph/0703583; Y. Gong and A. Wang, arXiv:0705.0996v1 [astro-ph]. 

\bibitem{Hawking} S. W. Hawking and G.F.R. Ellis, {\em The Large Scale Structure of Spacetime},(Cambridge University Press, England, 1973).

\bibitem{Barrow-Cotsakis-Magnano} J.D. Barrow and S. Cotsakis, Phys. Lett. B 
\textbf{214}, 515 (1988). See also G. Magnano and L. Sokolowski, Phys. Rev. D {\bf 50}, (1994) 5039; D. Wands, Class. Quantum Grav. {\bf 11}, (1994) 269; S. Baghram, M. Farhang and S. Rahvar, Phys. Rev. D {\bf 75}, (2007) 044024.

\bibitem{Kar-Dadhich} S. Kar and S. SenGupta, arXiv:gr-qc/0611123v1 (2006);
N.Dadhich, arXiv:gr-qc/0511123. 

\bibitem{Hwang} J.C. Hwang, Classical Quantum Gravity {\bf 7}, (1990) 1613; S. Capozziello, Int. Journ. Mod. Phys. D {\bf 11}, (2002) 483;  S. Capozziello, V.F. Cardone, S. Carloni and A. Troisi, Int. J. Mod. Phys. D {\bf 12}, (2003) 1969; T. Multam$\ddot{a}$ki and I. Vilja, Phys. Rev. D {\bf 73}, (2006) 024018;  G.J. Olmo, Phys. Rev. D {\bf 75}, (2007) 023511 and gr-qc/0612047.


\bibitem{Barrow-Ottewill} J.D. Barrow and A.C. Ottewill, J. Phys. A:
Math. Gen. \textbf{16}, 2757 (1983).

\bibitem{Barrow2004} J.D. Barrow, Class. Quantum Grav. {\bf 21}, L79
(2004).

\bibitem{Barrow2004a} J.D. Barrow, Class. Quantum Grav. {\bf 21},
5619 (2004).

\bibitem{Stefancic} H. \v{S}tefan\v{c}i\'{c}, Phys. Lett. B {\bf 586}
(2004), 5; {\it ibidem} B{\bf 595} (2004), 9; Phys. Rev. D{\bf 71}
(2005), 084024.

\bibitem{Nojiri} S. Nojiri and S.D. Odintsov, Phys. Lett. B {\bf 595} (2004), 1;
S. Nojiri, S.D. Odintsov and S. Tsujikawa, Phys. Rev. D {\bf 71} (2005), 063004.

\bibitem{Dabrowski} M.P. D\c{a}browski Phys. Lett. B \textbf{625}, 184 (2005).

\bibitem{Lazkoz04} L. Fernandez-Jambrina and R. Lazkoz, Phys. Rev.
D{\bf 70}, 121503 (2004).

\bibitem{Barrow-Tsagas2005} J.D. Barrow and C.G. Tsagas,
Class. Quantum Grav. {\bf 22}, 1563 (2005).


\bibitem{Visser} M. Visser, Class. Quantum Grav. {\bf 21}, (2004) 2603; Gen. Relativ. Gravit. {\bf 37}, (2005) 1541; E.R. Harrison, Nature \textbf{260}, 591 (1976); P. Landsberg, 
Nature \textbf{263}, 217 (1976).

\bibitem{Rapetti} D. Rapetti, S.W. Allen, M.A. Amin and R.D. Blandford, Mont. Not. R. Soc. {\bf 375}, (2007) 1510.

\bibitem{Turner} S.M. Carroll, A. De Felice, V. Duvvuri, D.A. Easson, M. Trodden and M.S. Turner, Phys. Rev. D {\bf 71}, (2005) 063513.

\bibitem{Sobouti} S. Capozziello, V.F. Cardone and A. Troisi, JCAP 08, (2006) 001;  Y. Sobouti, Astronomy \& Astrophysics {\bf 464}, (2007) 921;  C. F. Martins and P. Salucci, arXiv:astro-ph/0703243 (2007).

\bibitem{Capo} S. Capozziello, V.F. Cardone and A. Troisi, MNRAS {\bf 375}, (2007) 1423.

\bibitem{Mendoza} S. Mendoza and Y.M. Rosas-Guevara, arXiv:astro-ph/0610390 (2006).

\bibitem{Carloni} S. Carloni, P. Dunsby, S. Capozziello and A. Troisi, Class. Quantum Grav. {\bf 22}, (2005) 4839; S. Carloni, P. K. S. Dunsby and A. Troisi, arXiv:0707.0106 [gr-qc] (2007).


\bibitem{Timothy_John} T. Clifton and J.D. Barrow, Phys. Rev. D {\bf 72},
(2005) 103005.


\end{thebibliography}
\end{document}